\documentclass[12pt,preprint]{aastex}
\usepackage{psfig,epsfig,amsfonts,amsmath,amssymb,graphicx,morefloats,appendix,tensor}
\usepackage{color}
\usepackage{natbib}
\usepackage{hyperref}
\begin{document}

\slugcomment{Accepted to ApJL: October 11, 2022}

\title{First Millimeter Flares Detected From $\epsilon $ Eridani with ALMA}

\author{Kiana Burton \altaffilmark{1,2,3}, Meredith A. MacGregor\altaffilmark{1}, Rachel A. Osten\altaffilmark{4,5}}

\altaffiltext{1}{Department of Astrophysical and Planetary Sciences, University of Colorado, 2000 Colorado Avenue, Boulder, CO 80309, USA}
\altaffiltext{2}{Department of Physics, Temple University, 925 N 12th St, Philadelphia, PA 19122, USA  }
\altaffiltext{3}{Maria Mitchell Observatory, 4 Vestal St, Nantucket, MA 02554, USA}
\altaffiltext{4}{Space Telescope Science Institute, 3700 San Martin Drive, Baltimore, MD 21218 USA}
\altaffiltext{5}{Center for Astrophysical Sciences, Department of Physics and Astronomy, Johns Hopkins University, 3400 North Charles St., \\Baltimore MD 21218 USA}

\begin{abstract}

We report the detection of three large millimeter flaring events from  the nearby Sun-like, $\epsilon$~Eridani, found in archival ALMA 12m and ACA observations at 1.33~mm taken from 2015 January 17--18 and 2016 October 24--November 23, respectively. This is the first time that flares have been detected from a Sun-like star at millimeter wavelengths.  The largest flare among our data was detected in the ALMA observations on 2015 January 17 from 20:09:10.4--21:02:49.3 (UTC) with a peak flux density of 28 $\pm$ 7 mJy and a duration of 9 sec. The peak brightness of the largest flare is $ 3.4 \pm 0.9 \times 10^{14}$ erg s$^{-1}$Hz$^{-1}$, a factor of $>50\times$ times brighter than the star's quiescent luminosity and $>10\times$ brighter than solar flares observed at comparable wavelengths. We find changes in the spectral index (F$_\nu\propto\nu^\alpha$) at the flare peak, with $\alpha$ = 1.81 $\pm$ 1.94 and a lower limit on the fractional linear polarization $|Q/I| =$ 0.08 $\pm$ 0.12.  This positive spectral index is more similar to millimeter solar flares, differing from M dwarf flares also detected at millimeter wavelengths that exhibit steeply negative spectral indices. 

\end{abstract}

\section{Introduction}
\label{sec:intro}

$\epsilon $ Eridani is a K2V star \citep{Keenan:1998} at a distance of 3.22~pc \citep{vanLeeuwen:2007} making it the closest Sun-like star to host a circumstellar debris disk and potential planetary system.  With an estimated age of 400--800~Myr \citep{Mamajek:2008}, it is often considered to be an analogue of our own Solar System at an earlier stage of its evolution. Observations of this system have largely focused on the inner and outer debris disks, since its proximity allows for detailed analyses of their radial and azimuthal structures \cite[e.g.,][]{Greaves:1998,Greaves:2005a,Greaves:2005b,Backman:2009,Greaves:2014,Lestrade:2015,MacGregor:2015,Booth:2017}.  Although debated due to stellar activity, radial velocity observations spanning 20~years have been used to provide evidence for a planet ($\epsilon $ Eridani b) with an orbital period of approximately 7~years and a semi-major axis of 3.4~AU \citep{Hatzes:2000,Anglada:2012,Mawet:2019}. 

Observations at millimeter to radio wavelengths provide direct constraints on the properties of accelerated particles during stellar flares.  Previous radio observations of $\epsilon$ Eridani in the 2--4~GHz range with the Jansky Very Large Array (VLA) led to a flare detection, but it is unclear whether the source is the star or the planet \citep{Bastian:2018}. More recent VLA observations in this same frequency range detected continuum emission but found no flares \citet{Suresh:2020}. No millimeter flaring studies have been previously carried out for this target.

FUV, EUV, and X-ray observations are critical to characterize the effects of stellar radiation on exoplanetary atmospheres.  These high energy photons can dissociate molecules and ionize atoms, both modifying and heating atmospheres. $\epsilon $ Eridani has been well-observed at high-energy wavelengths. It was a target of the MUSCLES Treasury survey, in which XMM-Newton and HST STIS observations were obtained \citep{Loyd:2016} and detected a large flare simultaneously at FUV and X-ray wavelengths \citep{Loyd:2018}.  Interestingly, the X-ray and FUV emission during this flare are inconsistent, with the X-ray event lasting several times as long as the FUV event. A 3-year X-ray activity cycle was also reported by \citep{Coffaro:2020} using XMM-Newton lightcurves, and \citep{Linsky:2014} obtained flux ratios in the 10--20, 20--30, and 30--40~nm bands using data from the Extreme Ultraviolet Explorer. 

Recent multi-wavelength analyses of bright flares from Proxima Centauri suggest that millimeter flaring emission is correlated with the FUV \citep{MacGregor:2021}. Detecting and analyzing a larger sample of millimeter flares will be critical to help us evaluate this potential correlation and better understand what constraints millimeter flaring emission can help us place on the radiation environment of exoplanets. Despite this, stellar flares in the millimeter regime have been largely unexplored. Previous detection of flares at millimeter wavelengths have mostly come from M dwarfs (see Table \ref{tab:tab1}) \citep{MacGregor:2018a,MacGregor:2020}, but have also been reported at $\sim100$~GHz from V773 Tau \citep{Massi:2006}, $\sigma$ Gem \citep{Brown&Brown:2006}, UX Ari \citep{Beasley&Bastian:1998}, and GMR-A \citep{Bower:2003}. \citep{Mairs:2019} also report a bright sub-millimeter flare in a binary T Tauri system.  Our detection of millimeter flares with ALMA is the first time flares have been reported at millimeter wavelengths from $\epsilon$ Eridani.  Taken together, these detections indicate that millimeter emission is likely a common feature of stellar flares. More detections of variable millimeter emission will confirm this, with all-sky millimeter surveys providing the best means to increase the detection of flares from stars of various spectral types. 

 In this paper, we analyze archival observations from the Atacama Large Millimeter/submillimeter Array (ALMA) full 12-m array and Atacama Compact Array (ACA) of $\epsilon $ Eridani and report the discovery of three millimeter flaring events. Section \ref{sec:observations} discusses the details of the archival observations. In Section \ref{sec:analysis}, we present our new analysis of these observations, and approach to searching for and characterizing time-variable emission.  In Section \ref{sec:disc}, we discuss the properties of the $\epsilon$ Eridani flares in the context of those seen from M dwarfs and the Sun, and consider the complications of disentangling stellar and dust components without time- and wavelength-resolved observations.

\section{Observations}
\label{sec:observations}

Given that it is the closest Sun-like star to Earth that hosts a tentative planet along with both inner and outer debris disks, the $\epsilon$ Eridani system has been a frequent target for ALMA.  Archival observations are available using Band 6 (center wavelength of 1.3~mm) from both the full 12-m array and ACA.  Although, all of these observations originally targeted the outer debris disk, the reanalysis presented here can still yield important constraints on the properties of the central star.  A short description of both data sets follows. All analysis was performed using the Common Astronomy Software Package of \texttt{CASA} \cite[version 6.0.0.27][]{McMullin:2007}.  Imaging made use of the \texttt{tclean} task.

\subsection{ACA Observations}

The $\epsilon$ Eridani system was observed by the ACA between 2015 October 24 and 2015 November 23 (2016.1.00803.S, PI: MacGregor) for a total of 5 hours on-source.  These observations were split up into five $\sim$60-minute SBs where 6.66~min on-source integrations were interleaved with observations of a phase calibrator J0336-1302. The bright quasars J0334-4008, J0423-0120, J0519-4546, and J2357-5311 were used for bandpass calibration, while J0423-013, Uranus, and Mars were used for absolute flux calibration. The weather was excellent, with a precipitable water vapor (pwv) of 1.8~mm throughout the observations.

The original goal of these observations was to map the outer debris disk with uniform resolution and sensitivity.  In order to do that, a 7-pointing mosaic imaging pattern was adopted -- one pointing at the stellar position and six pointings spaced evenly around the disk. Figure~\ref{fig:fig1} (left panel) shows the natural weight continuum image obtained by combining all of these observations into a single map of the system. The rms noise is 94~$\mu$Jy~beam$^{-1}$.  The outer ring is clearly detected with an unresolved $10\sigma$ central point source at the stellar position.  Fitting a point source model to the combined visibilities yields a flux of $776 \pm 103$~$\mu$Jy.

\subsection{ALMA 12-m Observations}

$\epsilon $ Eridani was also observed with the full ALMA 12-m array with 34--35 antennas between 2015 January 17--18 (2013.1.00645.S, PI: Jord\'{a}n) for a total of 4.4 hours on-source. These observations were split into six $\sim$ 50 min scheduling blocks (SBs) with 7.66 min on-source integrations alternating with a phase calibrator J0336-1302. The bright quasars J0334-4008, J0423-0120, J0519-4546, and J2357-5311 were used for bandpass calibration, while J0423-013, Uranus, and Mars were used for absolute flux calibration. The weather was good overall, with the pwv ranging from 4.0--4.2~mm. Additional details about these observations can be found in \cite{Booth:2017}, who analyzed the northern arc of $\epsilon$ Eridani’s outer debris disk.

\begin{figure}[t]
  \begin{center}
       \includegraphics[scale=0.8]{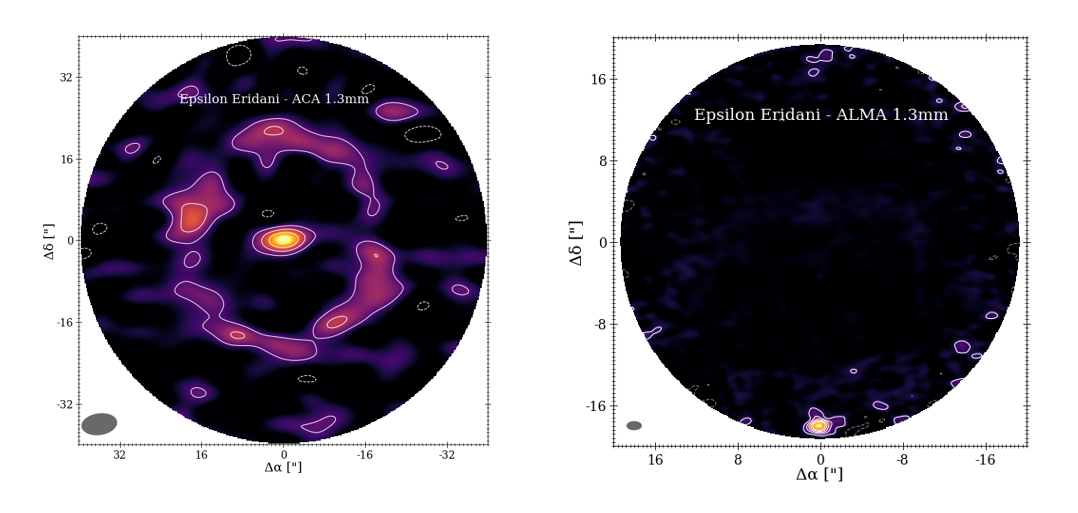}
  \end{center}
\caption{\small \textit{(left)} Natural weighted image of continuum emission from the ACA. Millimeter emission is detected at the stellar position (center of the image) at  $776 \pm 103$~$\mu$Jy . The outer debris disk is seen surrounding the star at approximately $16\arcsec$. The contour levels are in steps of $[-2,2,4,6,8...]\times$ the rms noise of 94 $\mu$Jy~beam$^{-1}$. The grey ellipse in the lower left corner shows the beam size which is $7\farcs4 \times 4\farcs5$ \textit{(right)} Primary beam corrected, natural weighted continuum image of $\epsilon$ Eridani and the northern arc of its outer debris disk from the ALMA 12-m array. Millimeter emission at the stellar position, $794\pm45$~$\mu$Jy. The contours are in steps of $[-3,3,6,9...]\times$ the rms noise of 33~$\mu$Jy~beam$^{-1}$ (primary beam corrected noise). The beam size is $1\farcs6\times1\farcs1$, as indicated by the grey ellipse in the lower left corner of the image.}
\label{fig:fig1}
\end{figure}

Figure~\ref{fig:fig1} (right panel) displays the primary beam corrected natural weight continuum image. The synthesized beam size is $1\farcs6\times1\farcs1$.  We determine the rms noise to be 33~$\mu$Jy~beam$^{-1}$, somewhat higher than the 14~$\mu$Jy~beam$^{-1}$ reported by \cite{Booth:2017}. {This discrepancy could stem from the fact that \cite{Booth:2017} determine the rms from the dirty image with no primary beam correction applied or be due to a different choice of regions when calculating the image statistics.  In order to calculate the rms, we choose ten regions that do not overlap with known emission sources (i.e., the star and surrounding disk), calculate the rms in each region, and average the values.}  The northern arc of the debris disk passes through the center of the image.  An unresolved $24\sigma$ point source is evident at the edge of the primary beam ($\sim18\arcsec$ from the pointing position) coincident with the expected stellar position.  We use the \texttt{uvmodelfit} task in \texttt{CASA} to fit a point source model to the visibilities and return a flux density of $794\pm45$~$\mu$Jy, consistent within the mutual uncertainties to the $820\pm70$~$\mu$Jy reported in \cite{Booth:2017}.  We note that this is somewhat in excess of the flux expected for the stellar photosphere alone. 

It is important to note that a primary beam correction has been applied to all of the images we present in this paper.  This function essentially divides the image by the antenna power response function in order to improve the accuracy of flux densities measured at locations off of the phase center.  Because the primary beam correction is applied uniformly, it not only `upweights' any real signals but also increases the apparent noise at the edges of the image (as can be seen in the right panel if Fig.~\ref{fig:fig1}) making it difficult to determine the uncertainty and judge significance levels.  As a result, we choose to fit the visibilities directly to determine the stellar flux and search for any flare candidates as is described below in Section~\ref{sec:analysis}.

\section{Results and Analysis}
\label{sec:analysis}

\subsection{Creating Light Curves and Identifying Flares}
\label{sec:lightcurve}

\begin{figure}[t]
  \begin{center}
       \includegraphics[scale=0.4]{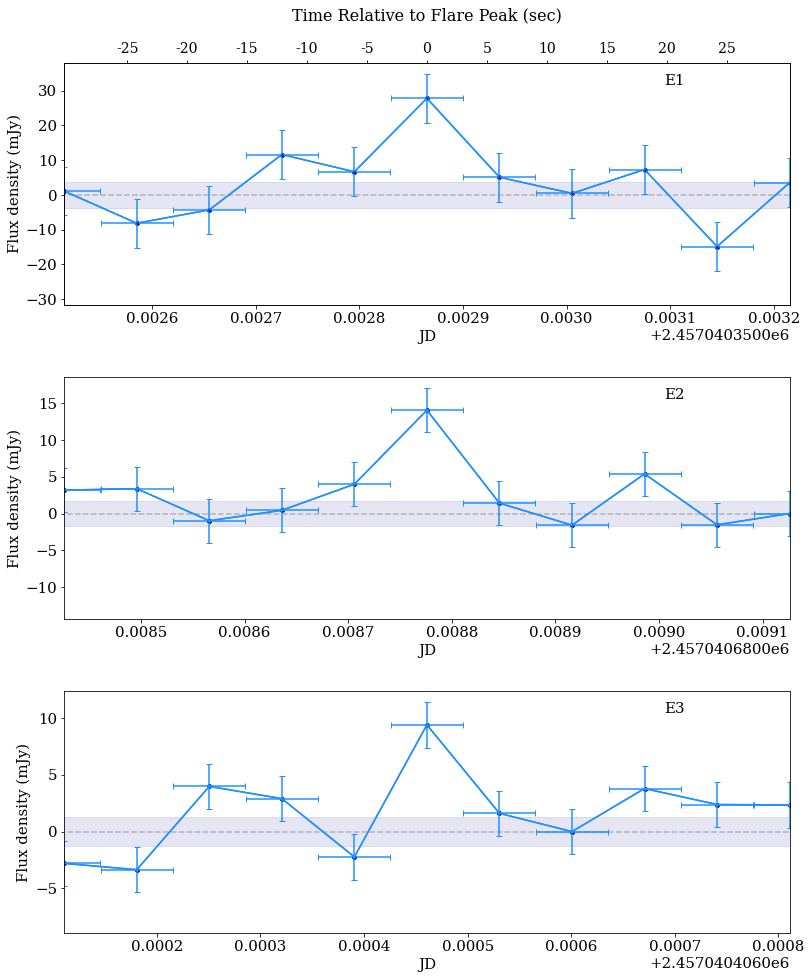}
  \end{center}
\caption{\small Light curves of the three major flaring events labeled E1, E2, and E3. Each light curve is centered on the flare peak, with the top x axis indicating time relative to the flare peak in seconds.  Each data point represents an individual ALMA integration (6~sec in length, indicated by the x error bars).  The shaded region in each plot indicates $\pm$ the rms noise level centered on a dashed line at 0~mJy.}
\label{fig:fig2}
\end{figure}

As in \cite{MacGregor:2018a} and \cite{MacGregor:2020}, we search for time-variable emission and flares by fitting point source models to the millimeter visibilities.  However, we have now developed an automatic pipeline approach that makes use of the \texttt{uvmodelfit} task in \texttt{CASA}.  For a given data set, models are fit to each integration from the telescope starting at the beginning of the observations and stepping forward in time.  The end product is essentially a light curve, with best-fit flux densities recorded for each integration.  For the ALMA 12-m and ACA observations, the integration times were 6 and 10~sec, respectively, setting the finest temporal resolution of the final light curves.  We note that there is a trade-off as the integration time shrinks -- Shorter integration times imply higher rms noise and fainter flares will not be detected.  However, millimeter flares have so far proven to be short events (typically lasting $<30$~sec), so fine time resolution is needed in order to fully characterize their temporal behavior.  Once the initial light curve is created, any recorded flux density that is above $3\sigma$ is flagged as a potential flaring event.  In order to confirm flares, we run a second pipeline and again fit point source models to the candidate flares this time to the lower (217 and 219~GHz) and upper (230 and 232~GHz) sidebands independently.  We require that the flux densities measured in both sidebands be within 20\% of each other.  Any events that do not meet this criteria are discarded as spurious noise.  

\begin{figure}[t]
  \begin{center}
       \includegraphics[scale=0.7]{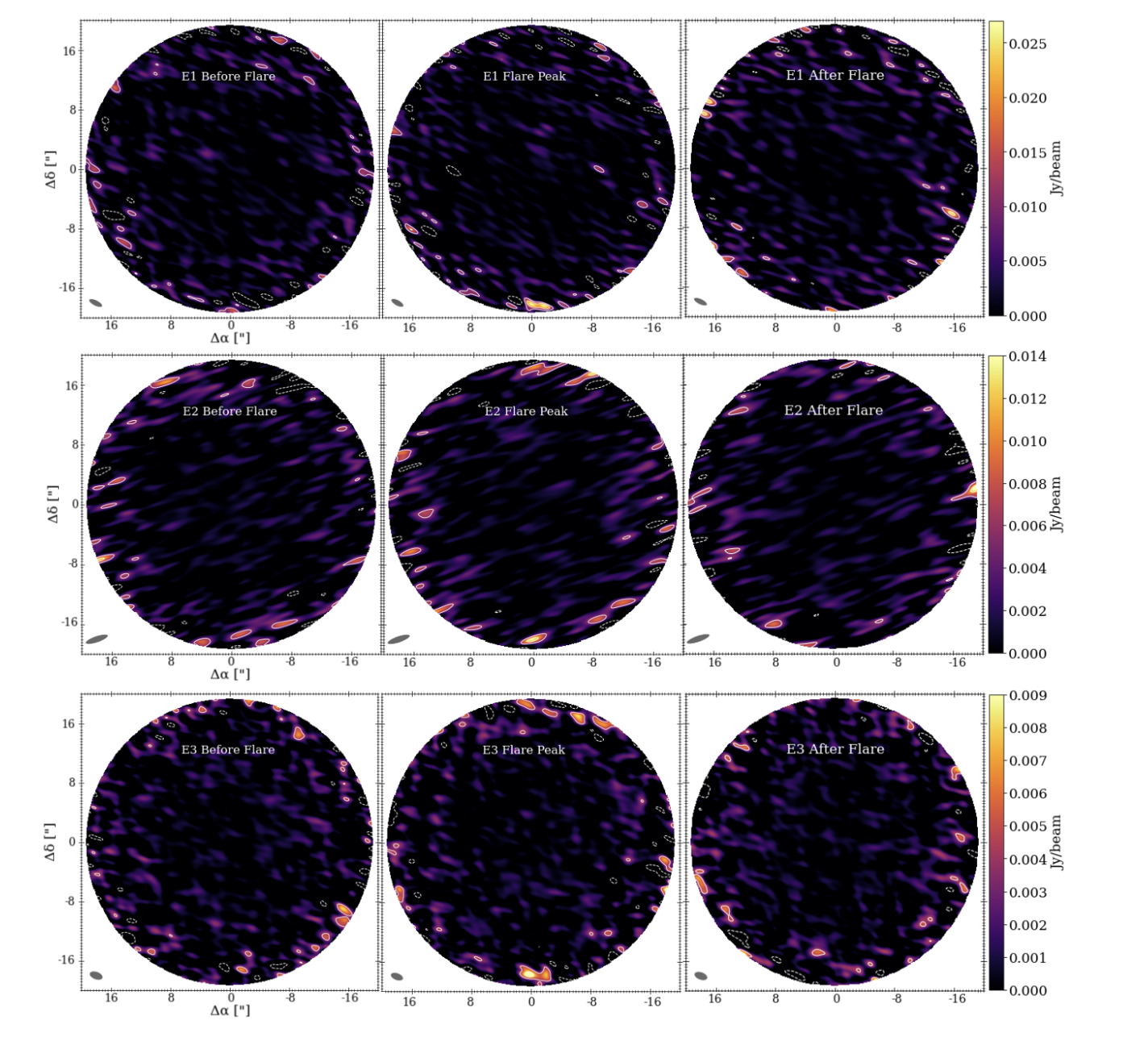}
  \end{center}
\caption{\small Images for all three detected flares (E1, top row; E2, middle row; E3, bottom row) in the integration just before the flare (left panels), at the peak of the flares (middle panels), and just after the flare (right panels).  A point source is only detected at the stellar position (bottom center of the image towards the edge of the primary beam) at the flare peak for all three events.  The flux density from these images (indicated by the color scale bars) are consistent with what we obtain from visibility modeling.}
\label{fig:fig3}
\end{figure}

For the ALMA 12-m observations, these pipelines revealed three large flaring events.  Each light curve is shown in Figure~\ref{fig:fig2}. The peak flux densities and luminosities for all three flares, along with other characteristics described in Section~\ref{sec:characteristics}, are reported in Table~\ref{tab:tab1}.  The brightest event (labeled E1, top panel of Figure~\ref{fig:fig2}) occurred on 2015 January 17 from 20:09:10.4--21:02:49.3 UTC.  During this flare, the star brightened by a factor of $50\times$ over quiescent values, reaching a peak flux density of $28\pm7$~mJy and peak luminosity of $3.4\pm 0.9 \times 10^{14}$~erg~s$^{-1}$~Hz$^{-1}$.  The other two flares reached lower peak flux densities of $14\pm 3$~mJy (labeled E2) and $9\pm 2$~mJy (labeled E3), corresponding to peak luminosities of $1.7\pm 0.4  \times 10^{14}$~erg~s$^{-1}$~Hz$^{-1}$ and $1.1\pm 0.2 \times 10^{14}$~erg~s$^{-1}$~Hz$^{-1}$, respectively.  We note that the uncertainty on the flux is higher for E1 since the weather was slightly worse at this time and thus the rms noise was larger. Figure~\ref{fig:fig3} shows images for each of these flaring events before (left), at peak (center), and after (right).  In the before and after images, no significant emission is detected at the stellar position, while at the flare peak a clear point source is visible for all three events.  The flux densities obtained from these images are consistent with those obtained from our visibility modeling, adding further credibility to these detections. 

No flares were found in the ACA observations.  However, with fewer antennas in the array and smaller antenna diameters (7-m for the ACA), the total collecting area is reduced and as a result the rms noise in each integration was a factor of $3-5\times$ higher than for the ALMA 12-m observations.  As a result, the three flares detected in the 12-m observations would not have been detected in these ACA observations.

\subsection{Characterizing Millimeter Flaring Emission}
\label{sec:characteristics}

\begin{figure}[t]
  \begin{center}
       \includegraphics[scale=0.35]{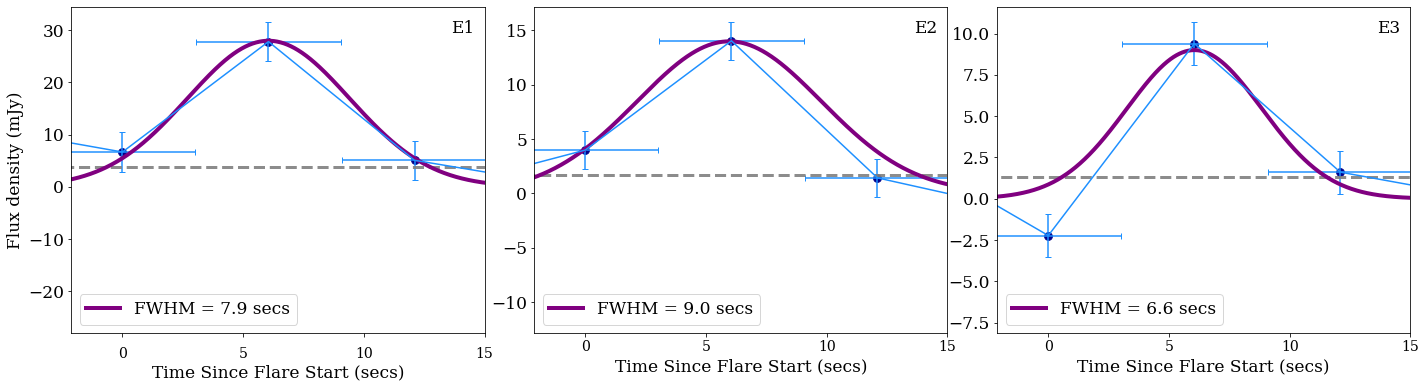}
  \end{center}
\caption{\small Gaussian fits to the light curves of the three flares detected from $\epsilon$ Eridani. These flares are similar to millimeter flares detected previously by ALMA in being relatively short-lived.}
\label{fig:fig4}
\end{figure}

To determine the temporal structure of the flaring events, we fit Gaussian functions to the light curves shown in Figure~\ref{fig:fig4}.  The $t_{1/2}$ or FWHM of the fits are shown on the plots in seconds. Event E3 has the shortest duration of only 6.6~sec, while the E1 and E2 flares do not last much longer, at 7.9 and 9.0~sec, respectively. The Gaussian fits reveal symmetric temporal structures and show that there is little exponential decay following the flare peak. This is similar to the structure of the millimeter flares reported from other M dwarfs. Likewise, Table~\ref{tab:tab1} shows that the AU Mic (labeled A) and Proxima Centauri (labeled P) flares also have very short durations. 

To examine the characteristics of the flaring events and consider emission mechanisms, we examine the spectral index (dependence of the flux on the frequency, F$_\nu\propto\nu^\alpha$) of the upper (230 + 232 GHz) and lower (219 + 217 GHz) sidebands and a lower limit on the fractional linear polarization ($|Q/I|$) for all three flaring events.  To obtain the spectral index, we fit point source models using \texttt{uvmodelfit} in \texttt{CASA} to the millimeter visibilities of the upper and lower sidebands independently. For the fractional linear polarization, we fit point sources models to ALMA's XX and YY polarizations independently, and compute the Stokes parameters $Q = <{E_{x}}^2>-<{E_{y}}^2> $ and $I = <{E_{x}}^2>+<{E_{y}}^2> $.  Due to limited signal-to-noise, we are only able to compute the spectral index and lower limit on the fractional linear polarization at the peak of the three detected flares.  Table 1 lists these values in comparison to all other millimeter flares previously detected with ALMA. 
We note that two flare events are consistent with no net linear polarization, and a third event (E2) is marginally significant.

\section{Discussion}
\label{sec:disc}

We have detected the first millimeter flares from a Sun-like star, $\epsilon$ Eridani, with ALMA.  As described above, we have characterized the temporal, spectral, and polarization characteristics of these three flaring events.  Below, we discuss the implications of these new detections for the population of millimeter stellar flares and the $\epsilon$ Eridani planetary system.  Section~\ref{sec:flares_disc} places the $\epsilon$ Eridani flares in context with other millimeter flares detected from our Sun and the M dwarfs Proxima Centauri and AU Mic.  In Section~\ref{sec:dust_disc}, we derive new constraints on the millimeter flux of any unresolved inner dust belts.

\subsection{Nature of the Detected Millimeter Flares}
\label{sec:flares_disc}

These new flares are detected from a K dwarf while all previous ALMA detections are from M dwarfs.  Still, their temporal characteristics appear  similar to previous flares detected by ALMA from Proxima Centauri and AU Mic, in being of relatively short duration.  All of the new events are short with $t_{1/2}$ values $<10$~sec, while previously detected millimeter flares have $t_{1/2}$ values of $2-35$~sec \citep{MacGregor:2018a,MacGregor:2020}. As a cumulative group, all of these events exhibit a roughly symmetric rise and fall with no indication of significant exponential decay as is typically seen for longer flares at other wavelengths \citep{Kowalski:2013,Kowalski:2016,Huenemoerder:2010}.  It is notable that all of the stars for which we have used ALMA to create millimeter light curves to date exhibit flaring emission \cite{MacGregor:2018a,MacGregor:2020, MacGregor:2021}.  In addition, all-sky cosmology surveys with the South Pole Telescope (SPT) and the Atacama Cosmology Telescope (ACT) have  produced a growing number of millimeter flaring detections \cite[e.g.,][]{Guns:2021,Naess:2021}.  Therefore, we expect that observations and analyses of stars at millimeter wavelengths with ALMA and other facilities will continue to yield many new detections of flares.

The new $\epsilon$ Eridani flaring events reported here are also similar to previous millimeter flares in that they show evidence for linear polarization at peak.  Although there are large uncertainties due to the low signal-to-noise ratio and lack of polarization calibration, the apparent polarization of millimeter flaring emission suggests that synchrotron \citep{Phillips:1996,Massi:2006} or gyrosynchrotron \citep{White:1992} emission is the likely emission mechanism for these events.  Most stellar sources of gyrosynchrotron emission display a dominant component of circular polarization \citep{Dulk:1985}, while the few stellar sources with convincing evidence of synchtrotron emission exhibit linear polarization \citep{Massi:2006}. 
Linear polarization can arise from propagation effects like Faraday rotation, which can convert circular to linear polarization. 
In \citet{Massi:2006} there was evidence not only for linear polarization, but extended coronal structures through the the polarization as well as long trapping timescales. 
Being both of short duration and evincing linear polarization at millimeter wavelengths is a new regime
of parameter space for stellar flares. 
This may be indicative of directly precipitating electrons, as suggested by
\citet{MacGregor:2020}. 
Another possibility could be 
an indication of differences in conditions for trapped electrons which might deviate from expectations \citep[for instance, trapping in the strong diffusion limit has a very different temporal dependence for energetic electrons in terms of number of accelerated particles at a given time compared to the weak diffusion limit,][]{Lee2002}.
Observational evidence is currently suggestive of the presence of accelerated particles in the atmosphere of $\epsilon$ Eridani, but more definitive proof of the emission mechanism (whether gyrosynchrotron or synchrotron) requires additional data and
modeling.

\citet{Valenti1995} used high resolution near-infrared spectra to determine the magnetic field strength of $\epsilon$ Eridani in the photosphere to consist of a 1.44~kG magnetic field covering 8.8\% of the surface.  More recent investigations of $\epsilon$ Eridani's magnetic field from Zeeman Dopppler Imaging of near-infrared and Fe~I lines \citep{Petit2021} suggest a slightly higher magnetic field strength, at 1.835~kG, and filling factor of 0.13. The physical region producing this flaring millimeter emission is likely associated with the chromosphere or corona above the photosphere.  Gyrosynchrotron emission is produced by a range of harmonic numbers $s$ of the electron gyrofrequency, $\nu_{B}=2.83\times10^{6} B (G)$, with $s$ between 10 and 100.  The properties of the emitting region under consideration here can be determined  generally  by associating the observing frequency with a harmonic of the gyrofrequency. This leads to an unbiased but large range of magnetic field strengths in the mm-emitting region of 0.8--8~kG.  The lower range of magnetic field strengths derived above is compatible with the photospheric value, suggesting an origin for the flare in a region close to the surface of the star.  
More quantitative constraints on the magnetic field strength in the millimeter-emitting region rely on models that can reproduce both the observed intensity as well as spectral index and polarization, and is beyond the scope of the current paper.  Synchrotron emission generally occurs for a range of harmonics and can extend up to large values of harmonic numbers \citep{RybickiLightman}.  In the current situation this would imply exceedingly large values of magnetic field strengths.


The one characteristic where these new $\epsilon$ Eridani events that differs from the previous M dwarf events is the spectral index.  All of the flares previously detected from AU Mic and Proxima Centauri exhibited steeply negative spectral indices at peak.  In our new sample, only event E3 shows a similar negative spectral index.  In contrast, events E1 and E2 exhibit positive $\alpha$ values.  Intriguingly, this is similar to what has been observed from solar flares at comparable wavelengths. \cite{Krucker:2013} report the detection of several solar millimeter flares with $\alpha = 0-6$. The brightest solar flared observed had a luminosity of  $\sim2\times10^{13}$~erg~s$^{-1}$~Hz$^{-1}$, while the largest flare detected from $\epsilon$ Eridani was more than $10\times$ brighter with a luminosity of $3.4\times10^{14}$~erg~s$^{-1}$~Hz$^{-1}$. Smaller solar flares reported in \cite{Krucker:2013} have luminosities between $1-5\times10^{12}$~erg~s$^{-1}$~Hz$^{-1}$, more than $100\times$ less luminous than any of the detected $\epsilon$~Eridani flares.  The similarity in observed spectral index at millimeter wavelengths between solar flares and these new $\epsilon$~Eridani flares in contrast to previously detected M dwarf millimeter flares indicates that there is more to learn by examining a larger population of flares from stars of differing spectral types.  Perhaps millimeter observations probe different regions of the synchrotron or gyrosynchrotron spectrum for different spectral type stars.

\subsection{Constraints on Dust Emission from an Inner Debris Disk}
\label{sec:dust_disc}

$\epsilon$ Eridani is the closest Sun-like star to Earth that hosts both a system of remnant debris disks and a potential planetary system.  As a young analogue to our Solar System, it has been the focus of many observational campaigns.  The outer Kuiper Belt analogue centered at $\sim60$~AU (seen in Figure~\ref{fig:fig1} $\sim18\arcsec$ from the stellar position) has been especially well-studied at infrared and millimeter wavelengths \cite[e.g.,][]{Greaves:1998,Greaves:2014,Lestrade:2015,MacGregor:2015,Chavez:2016,Booth:2017}.  An inner warm component several AU from the star has been inferred from infrared observations \cite[e.g.,][]{Backman:2009,Greaves:2014}.  Recently, \cite{Su:2017} marginally resolved this inner component through a combination of SOFIA and Spitzer images at 35 and 24~$\mu$m, respectively.  

Previous millimeter imaging studies noted excess emission from the central point source, presumed to be a combination of stellar and dust emission.  The effective temperature of $\epsilon$~Eridani is $5039\pm126$~K \citep{Baines:2012}, which implies a photospheric flux at 1.3~mm of $0.53$~mJy \cite[with 2\% uncertainty, see][]{Backman:2009}.  \cite{Lestrade:2015} and \cite{MacGregor:2015} determined flux densities of $1.2\pm0.3$ and $1.08^{+0.19}_{-0.41}$~mJy, respectively, at 1.3~mm, both significantly in excess of the expected stellar photosphere.  \cite{MacGregor:2015} also noted a flux excess at 7~mm with ATCA, which implied an increasing effective temperature with increasing wavelength, characteristic of stellar chromospheric emission as has been observed previously from $\alpha$ Cen A and B \citep{Liseau:2015,Liseau:2016}. \cite{Mohan:2021} fit a PHOENIX stellar atmosphere model to the available long wavlength data with a temperature of 5100~K.  

Our new detection of millimeter flares from $\epsilon$~Eridani adds another piece to the puzzle.  If these flares are truly common occurrences, they may have contributed to previous determinations of excess emission at comparable wavelengths.  We use \texttt{uvmodelfit} \texttt{CASA} to determine the flux density of the star over all of the available ALMA observations, excluding the time when the star was flaring (i.e., during events E1, E2, and E3).  The result is $787\pm45$~$\mu$Jy, less than previous millimeter measurement but still in excess of the expected stellar photosphere by $257\pm34$~$\mu$Jy.  The beam size in the 12-m ALMA observations is $\sim1\farcs6$ or $\sim5$~AU.  The potential inner dust belt(s) have been proposed to extend from either 3--21~AU \citep{Greaves:2014} or 1.5--2~AU and 8--20~AU \citep{Backman:2009}.  In either of these two models, the inner dust would not be resolved from the central star by the existing ALMA observations.  Therefore, we can assume that our flux measurement and all previous millimeter flux measurements contain some contribution from the star (including both temporally varying flare emission and more constant chromospheric emission) and some contribution for dust.  Given that the previous millimeter measurements are much in excess of what we report here, those values likely include significant flaring emission.  Extrapolating the SED fits for the inner dust ring from \cite{Su:2017} yields a predicted flux at 1.3~mm of $\sim0.2-0.3$~mJy, strikingly consistent with what we determine here.  We emphasize that other researchers should exercise caution when interpreting excess unresolved emission at millimeter wavelengths from $\epsilon$~Eridani and other targets.  Completely disentangling stellar and dust components will likely require time- and wavelength-resolved observations.

\section{Conclusions}
\label{sec:conc}
Our findings suggest that the millimeter flaring properties of $\epsilon$ Eridani appear to be more similar to those of the Sun than to those of M dwarf stars. This could indicate that there is some variation in millimeter flare properties from stars of different spectral types. Because millimeter flaring emission now appears to be a common part of stellar flaring that can enhance our understanding of flare physics especially particle acceleration, our work opens up new observational windows for understanding stellar flares from Sun-like stars. Future work will require additional detections of millimeter flares from Sun-like stars with high signal to noise so that more definitive conclusions can be drawn regarding flare properties. We also stress the importance of time- and wavelength-resolved observation to disentangle stellar and dust components. 

\vspace{1cm}
KB acknowledges support from the Maria Mitchell Observatory summer REU program funded through the National Science Foundation (NSF).  M.A.M. acknowledges support for part of this research from the National Aeronautics and Space Administration (NASA) under award number 19-ICAR19\_2-0041.  This paper makes use of the following ALMA data: ADS/JAO.ALMA \#2013.1.00645.S and \#2016.1.00803.S. ALMA is a partnership of ESO (representing its member states), NSF (USA) and NINS (Japan), together with NRC (Canada) and NSC and ASIAA (Taiwan) and KASI (Republic of Korea), in cooperation with the Republic of Chile. The Joint ALMA Observatory is operated by ESO, AUI/NRAO and NAOJ. The National Radio Astronomy Observatory is a facility of the National Science Foundation operated under cooperative agreement by Associated Universities, Inc.

\software{\texttt{CASA} \cite[6.0.0.27,][]{McMullin:2007}, \texttt{astropy} \citep{astropy:2013,astropy:2018}}


\bibliography{References.bib}


 \begin{deluxetable}{l c c c c c c c}
\tablecolumns{7}
\tabcolsep0.1in\footnotesize
\tabletypesize{\small}
\tablewidth{0pt}
\tablecaption{Millimeter Properties for All Detected ALMA Flares \label{tab:tab1}}
\tablehead{
\colhead{Star} &
\colhead{Flare$^{\dagger}$} & 
\colhead{Peak Flux Density} &
\colhead{Peak $L_R$} & 
\colhead{t$_{1/2}$} &
\colhead{$\alpha$} &
\colhead{$|Q/I|$} &\\
\colhead{} &
\colhead{} & 
\colhead{(mJy)} &
\colhead{$10^{13} $erg $s^{-1}$ Hz$^{-1}$} & 
\colhead{(sec)} &
\colhead{} &
\colhead{} &
}
\startdata
\centering
AU Mic & A1 & 15 & 196 & 35 & -1.30 $\pm$0.05 & $>$0.12$\pm$0.04 \\
       & A2 & 5 & 69 & 9 & $\ddag$ & $\ddag$ \\
\hline
Proxima Cen & P1 & 45 & 9.2 & 4 & $\ddag$ & $\ddag$\\
             & P2 & 20 & 4.1 & 2.8 & $\ddag$ & $\ddag$\\
             & P3 & 10 & 2.0 & 2.4$^\text{a}$ & $\ddag$ & $\ddag$\\
             & P4 & 100 & 20 & 16.4 & -1.77 $\pm$ 0.45 & $>$0.19 $\pm$0.02\\
             & P5 & 106 & 21 & 2.8 & -2.29 $\pm$ 0.48 & $>$-0.19 $\pm$0.07\\
 \hline
$\epsilon$ Eridani & E1 & 28 & 34 & 7.9$^\text{a}$ & 1.81 $\pm$ 1.94 & $>$0.08$\pm$0.12\\ 
& E2 & 14 & 17 & 9.0$^\text{a}$  & 7.29 $\pm$ 2.89 & $>$-0.48 $\pm$ 0.15 \\  
& E3 & 9 &  11 & 6.6$^\text{a}$ & -2.83 $\pm$2.33 & $>$-0.11$\pm$0.19  \\ 
\enddata
$^\text{a}$These flares are not well resolved temporally given the integration time of ALMA and their true durations could be shorter. The $\chi^2$ values resulting from the Gaussian fits to the $\epsilon$ Eridani flares are 0.012 (E1), 0.23 (E2), and 17.61 (E3).
\end{deluxetable}

\end{document}